\documentclass{emulateapj}
\usepackage{apjfonts,graphics}

\newcommand{\simgt}{\lower.5ex\hbox{$\; \buildrel > \over \sim \;$}}
\newcommand{\simlt}{\lower.5ex\hbox{$\; \buildrel < \over \sim \;$}}

\newcommand{\baredth}{\;\overline{\raise1.0pt\hbox{$'$}\hskip-6pt \partial}\;}
\newcommand{\edth}{\;\raise1.0pt\hbox{$'$}\hskip-6pt\partial\;}

\usepackage{epsfig}
\usepackage{subfigure}
 
\begin{document}

\title{Comparison of Cluster Lensing Profiles with 
$\Lambda$CDM Predictions\altaffilmark{1}}


\author{Tom Broadhurst\altaffilmark{2}, Keiichi Umetsu\altaffilmark{3,4},
Elinor Medezinski\altaffilmark{2}, Masamune Oguri\altaffilmark{5},
Yoel Rephaeli\altaffilmark{2,6}}

\altaffiltext{1}{Based on data collected at Subaru Telescope, which is
operated by the National Astronomical Observatory of Japan.} 
\altaffiltext{2}{School of Physics and Astronomy Tel Aviv University, Israel}
\altaffiltext{3}{Institute of Astronomy and Astrophysics, Academia Sinica,  P.~O. Box
23-141, Taipei 10617,  Taiwan}
\altaffiltext{4}{Leung Center for Cosmology and Particle Astrophysics,
      National Taiwan University, Taipei 10617, Taiwan}
\altaffiltext{5}{Kavli Institute for Particle Astrophysics and
Cosmology, Stanford University, 2575 Sand Hill Road, Menlo Park, CA
94025, USA} 
\altaffiltext{6}{Center for Astrophysics and Space Sciences, UC San Diego, 
California, USA}

\begin{abstract}

We derive lens distortion and magnification profiles of four well
known clusters observed with Subaru. Each cluster is very well fitted
by the general form predicted for Cold Dark Matter (CDM) dominated
halos, with good consistency found between the independent distortion
and magnification measurements. The inferred level of mass
concentration is surprisingly high, $8<c_{\rm vir}<15$ ($\langle
c_{\rm vir}\rangle = 10.39 \pm 0.91$), compared to the relatively
shallow profiles predicted by the $\Lambda$CDM model, $c_{\rm
vir}=5.06 \pm 1.10$ (for $\langle M_{\rm vir}\rangle =1.25\times
10^{15}M_{\odot}/h)$. This represents a $4\sigma$ discrepancy, and
includes the relatively modest effects of projection bias and profile
evolution derived from N-body simulations, which oppose each other
with little residual effect. In the context of CDM based cosmologies,
this discrepancy implies clusters collapse earlier ($z\geq 1$) than
predicted ($z<0.5$), when the Universe was correspondingly denser.

\end{abstract}                   


\section{Introduction}\label{section1}

Dark matter (DM) is understood to comprise $\simeq 85\%$ of the
material Universe (Fukugita \& Peebles 2004), in the form of massive
halos around galaxies and clusters of galaxies. The density profile of
a dark halo depends on the unknown nature of the DM and the way
structure develops over cosmic time. Galaxy halos may be modified
significantly when gas associated with the disk of a galaxy cools and
condenses during its formation. In contrast, clusters are so massive
that the virial temperature of the gas (typically, $3-15$ keV) is too
high for efficient cooling and hence the cluster potential reflects
the dominant DM.

Models for the development of structure now rest on accurate
measurements of the power spectrum of mass fluctuations, the
cosmological density of DM, and the contribution of a cosmological
constant, $\Lambda$ (Spergel et al. 2007; Komatsu et al. 2008). This
framework is the standard ``$\Lambda$CDM'' cosmological model, with
the added simple assumptions that DM is collisionless, reacts only
gravitationally, was never relativistic ('Cold'), and with initially
Gaussian-distributed density perturbations. In this context, detailed
$N$-body simulations have established a clear prediction that
CDM-dominated cluster halos should have relatively shallow,
low-concentration mass profiles, where the logarithmic gradient
flattens continuously toward the center with a central slope tending
towards $r^{-1}$, interior to a characteristic radius, $r_s\lesssim
100-200 {\rm kpc}\, h^{-1}$ (Navarro, Frenk, \& White 1997, NFW, hereafter;
Bullock et al. 2001; Maccio et al 2006; Hennawi et al. 2007; Neto et
al. 2007; Duffy et al. 2008).  This ``NFW'' profile is characterized
by the (total) mass, $M_{\rm vir}$, within the virial radius, $r_{\rm
vir}$, and by the concentration parameter, $c_{\rm vir}\equiv r_{\rm
vir}/r_s$.

Interaction between clusters indicates DM is collisionless, in
particular the ``bullet cluster''(Markevitch et al. 2002) - where
shocked gas lies between two substantial galaxy clusters, implying
these clusters passed through each other recently (Markevitch et
al. 2002). Here the weak lensing signal follows the bimodal
distribution of member galaxies reflecting substantial DM associated
with each cluster component (Clowe et al. 2006) and that this DM is
relatively collisionless like the galaxies. These observations
disfavour the class of alternative gravity theories for which the
lensing signal is expected to trace the dominant baryonic contribution
of the gas (Clowe et al. 2006). Other cases of interaction show that in
general displacement of the gas relative to the DM is typically
related to interaction (Jee et al. 2006; Okabe \& Umetsu 2008).
 
For many clusters no obvious evidence of recent interaction is seen as
the gas and member galaxies follow a symmetric, structureless
distribution. Measurements of the mass profiles of these relaxed
clusters may help in understanding the nature of DM, preferably
relying on gravitational lensing signals where model-dependent
assumptions are not required. For A1689, over 100 multiply-lensed
images have been used to derive the inner mass profile (Broadhurst et
al. 2005a), with the outer profile determined from weak lensing
(Broadhurst et al. 2005b). Together, the full profile has the predicted
NFW form, but with a surprisingly high concentration when compared to
the shallow profiles of the standard $\Lambda$CDM model (Broadhurst et
al. 2005b; Umetsu \& Broadhurst 2008). Furthermore, good consistency is
also found between the lensing based mass profile and the X-ray and
dynamical structure of this cluster in model independent analyses
(Lemze et al. 2007; Lemze et al. 2008, in preparation). A similar
discrepancy is also indicated by lensing observations of other
clusters e.g., MS2137-23 (Gavazzi et al. 2003) and CL0024+16 (Kneib et
al. 2003).

Here we examine a sample of relaxed, high-mass clusters, to test the
distinctive prediction of $\Lambda$CDM, and to settle empirically
whether the mass profile of A1689 is unusual. In section \S1 we
describe the data reduction in \S3 we present the weak lensing
analysis and the magnification profiles derived from background number
counts and make comparison with the prediction of the $\Lambda$CDM;
in \S4 we discuss our results and conclusions.

\section{Data Reduction} 

\begin{deluxetable*}{ccccccccc}
  \tablecaption{ the Subaru distortion measurements
    combined with the Einstein-radius constraint.}
  \tabletypesize{\scriptsize}  
  \tablewidth{\textwidth} 
  \tablehead{ 
    \colhead{Cluster} & 
    \colhead{$z$} & 
    \colhead{Filters} & 
    \colhead{Einstein Radius} & 
    \colhead{$\langle D_{ds}/D_s\rangle$} & 
    \colhead{$\frac{d\log N(<m)}{dm}$} &  
    \colhead{$M_{\rm vir}$} & 
    \colhead{ $c_{\rm vir}$} & 
    \colhead{$\chi^2/{\rm dof}$} \\
    \colhead{} &
    \colhead{} & 
    \colhead{} &
    \colhead{($\arcsec$)} &
    \colhead{} &
    \colhead{} &
    \colhead{($10^{15}M_\odot/h_{70}$)} &
    \colhead{} & 
    \colhead{} 
}
\startdata
  A1689 & 0.183 & $V_{\rm J}i'$ & 52 ($z_s=3.05$) & $0.704$ & 0.150 &
  $1.59^{+0.24}_{-0.22}$ &  $15.69^{+3.96}_{-2.88}$  & $4.94/9$ \\
  A1703 & 0.258 & $g'r'i'$ & 33 ($z_s=2.8$) & $0.722$ & 0.062 &
  $1.30^{+0.24}_{-0.20}$ & $9.92^{+2.39}_{-1.63}$    & $2.69/5$ \\
  A370 & 0.375 & $BR_{\rm C}z'$ & 43 ($z_s=1.5$) & $0.606$ & 0.088 &
  $2.93^{+0.36}_{-0.32}$ & $7.75^{+1.12}_{-0.92}$    & $5.54/8$ \\
  RX J1347-11 & 0.451 & $V_{\rm J}R_{\rm C}z'$ & 35 ($z_s=1.8$) & $0.553$ & 0.066 &
  $1.47^{+0.26}_{-0.23}$ & $10.42^{+3.25}_{-2.13}$    & $6.25/7$ \\
  \enddata

\end{deluxetable*}

We analyse deep images of A1689, A1703, A370, RXJ1347-11 taken by
the wide-field camera Suprime-Cam ($34\prime\times 27\prime$, Miyazaki
et al. 2002) in Subaru (8.3m) which are observed deeply in several
optical passbands, listed in Table~1. These clusters are of interest
due to the exceptional quality of the data, with exposures in the
range 2000s-10000s per pass band, with seeing ranging from 0\arcsec.5
to 0\arcsec.75. We use either the R or I bands for our weak lensing
measurements (described below \S3) for which the instrumental
response, sky background and seeing conspire to provide the best
images. All the available bands are used to define colours with which
we separate cluster members from the background in order to minimise
dilution of the weak lensing signal by unlensed objects.


 The standard pipeline reduction software for Suprime-Cam (Yagi et
 al. 2002) is applied for flat-fielding, instrumental distortion
 correction, differential refraction, PSF matching, sky subtraction
 and stacking. Photometric catalogs are constructed from stacked and
 matched images using Sextractor (Bertin \& Arnaut 1996). We select
 red galaxies with colours redder than the colour-magnitude sequence
 of cluster E/SO galaxies. The sequence forms a well defined line due
 to the richness and relatively low redshifts of our clusters. These
 red galaxies are expected to lie in the background by virtue of
 k-corrections which are greater than for the red cluster sequence
 galaxies and convincingly demonstrated spectroscopically by Rines \&
 Geller (2008). We also include very blue galaxies falling far from
 the cluster sequence,$> 1.^{m}5-2.^{m}5$, depending on the passbands
 available for each cluster (listed in Table 1), to minimise cluster
 contamination (see Medzinski et al. 2007). Typically the proportion
 of blue galaxies used is around 50\% of the red background. In
 addition, we adopt a conservative magnitude limit of $m<25.5-26.0$,
 depending on the depth of the data for each cluster, to avoid
 incompleteness. Imposing these strict limits is a necessary
 precaution against the diluting effect that unlensed cluster and
 foreground galaxies will otherwise produce, leading to spuriously
 shallow central mass profiles, as shown in Broadhurst et al. 2005b and
 Medezinski et al. 2008.

\section{Measurements of Lensing Effects}

These data permit high quality weak lensing measurements, following
established methods required to deal with instrumental and atmospheric
effects following the formalism outlined in Kaiser, Squires, \&
Broadhurst (1995), with modifications described in Erben et
al. (2001). The mean residual stellar ellipticity after PSF correction
is less than, $\sim 10^{-4}$, in all fields, averaged over 400-600
stars and consistent with the standard error on this
measurement, $\simeq 10^{-4}$. Note also the lack of any systematic
deviation from zero in the B-mode, $g_\times$, shown in Figure
2a. Full details of the methods are presented in Umetsu \& Broadhurst
(2008).

Figure 1 shows the gravitational shear field by locally averaging
the corrected distortions of colour-selected background galaxies of
each cluster. This is compared to the distribution of color-selected
cluster sequence galaxies. In each case, one large symmetric cluster
is visible around which the lensing distortion pattern is clearly
tangential, with little significant substructure. The derived radial
profiles of the lensing distortion are seen to be very similar in form
(left panel of Figure 2), with differing amplitudes reflecting a range
of mass. 

The NFW profile fits well each cluster (Figure 2; Table 1), but
surprisingly the derived concentrations lie well in excess of the
standard $\Lambda$CDM model (Figure 3). For the derived masses of our
clusters, the predicted mean concentration is $c_{\rm vir}\sim 5.06\pm
1.1$ based on Duffy et al. (2007) using the improved WMAP5 (Komatsu et
al.  2008). Our best-fitting values instead range over $8<c_{\rm
vir}<15$, with a mean value of ($\langle c_{\rm vir}\rangle = 10.39
\pm 0.91$) representing a $4.8\sigma$ discrepancy. Our measurements
also imply A370 is the most massive cluster now known, $M_{\rm
vir}=2.93 \times 10^{15}M_{\odot}/h_{70}$, with a virial mass twice
that of RXJ1347-11, the most X-ray luminous cluster known. Our
estimate of the virial mass of RXJ1347-11 is in very good agreement
with independent X-ray and lensing analyses (Miranda et al. 2007,
Kling et al. 2005). For A1689, a somewhat lower concentration parameter
is derived by Limousin et al. (2007), $c_{\rm vir}=9.6\pm 2.0$, from
independent weak lensing measurements, which is consistent with our
findings at the $1.5\sigma$ level, whereas the observed Einstein
radius of approximately 52\arcsec is under-predicted by the
NFW parameters obtained by Limousin et al. (2007), $24\arcsec
\pm 11\arcsec$ (see Umetsu \& Broadhurst 2008).

Note, when model fitting an estimate of the background depth is
required. The mean depth is sufficient for our purposes as the
variation of the lens distance ratio, $D_{ds}/D_s$, is slow for our
sample because the clusters are at relatively low redshift compared to
the redshift range of the background galaxies.  The estimated mean
depth of the combined red+blue background galaxies is listed in column
5, which we obtain by applying our colour-magnitude selection to
Subaru imaging of the HDF region (Capak et al. 2004), where photometric
redshifts are reliable for red galaxies, with a mean redshift close to
$z=1.2$ depending on the details of the colour magnitude
selection. For the blue galaxies we may rely on the zCOSMOS deep
redshift survey (Lilly et al. 2008) where our mean depth is typically
$z\simeq 2.1$.

Lensing measures projected mass and so a statistical bias arises from
the triaxiality of clusters in cases where the major axis lies along
the line-of-sight. This leads to an $\sim 18\%$ increase in the mean
value of lensing derived concentrations based on $\Lambda$CDM (Oguri
et al. 2005; Hennawi et al. 2007), and an overall discrepancy in $c_{\rm
vir}$ of $4.0\sigma$ with respect to the predictions. A larger bias
is inferred for clusters selected by the presence of large arcs, $\sim
34\%$, representing the most triaxial cases (Hennawi et al. 2007). Our
sample is defined by the quality of available imaging and includes
clusters observed for reasons other than lensing, hence it is unlikely
that these clusters are all particularly triaxial with the long axes
pointing to the observer.  Even so, applying the maximum estimated
bias ($\sim50\%$, Oguri \& Blandford 2008) cannot account
for our measurements (Figure 3).

Multiply-lensed images are visible in all our clusters, from which the
inner mass distribution may be determined (Broadhurst et al. 2005a;
Halkola et al. 2008; Hennawi et al. 2008) and an equivalent Einstein
radius derived by averaging azimuthally (Figure 2) which for most
cases is close to the observed radius. For A1703 and A370, the
tangential critical curves are significantly elongated, for
$r<1\arcmin$ (Limousin et al.  2008; Kneib et al. 1993) although in
general the mass distribution is always less elliptical than the
critical curve. X-ray observations of A370 show the inner region has
some substructure (eg. Shu etal 2008), but beyond $r>1\arcmin$ the
weak lensing pattern and the distribution of galaxies is 
symmetrical, as seen in Figure 1. For each cluster the NFW parameters
derived with or without the Einstein radius constraint are found to be
closely consistent and in combination they are more precisely
determined - see Table 1.

We also examine the magnification profile, $\mu(r)$, via the surface
number density of background galaxies (Broadhurst, Taylor, \& Peacock
1995). At faint fluxes where the counts follow a power-law slope,
$s=d\log N(<m)/dm$, lensing modifies the true density above the
magnitude limit , $N_0(m<m_{\rm lim})$, by, $N(<m_{\rm
lim})=N_0(<m_{\rm lim})\mu(r)^{2.5s-1}$, implying competition between
the magnified sky, which reduces the surface density, and an increase
of galaxies magnified above the flux limit.  Here we use only galaxies
lying redward of the cluster sequence because for these the intrinsic
count-slope of faint red galaxies is relatively flat, $s \sim 0.1$, so
a net count depletion results, which we readily detect for each
cluster, increasing towards the center (Figure 2). For each cluster we
calculate the depletion of the counts expected for the the best
fitting NFW profile derived from the corresponding distortion
measurements above (Figure 2, right panel, with same colour code),
finding clear consistency, which considerably strengthens our
conclusions, and also establishes the utility of the
background red galaxies for measuring magnification. The variation in
the total number of background counts seen in Figure 2 simply reflects
the relative depth of the imaging data and also to some extent the
redshift of the cluster, which lies redder in colour space with
increasing cluster redshift, thereby reducing the numbers of
background galaxies which can be selected to lie redward of the
sequence for our purposes. Note, absolute counts are not used here,
the models are instead normalised to the observed density when
comparing with the model to avoid reliance on control fields.

\section{Discussion}

We have shown that the mass distribution of a sample of the most
massive clusters follow a very similar form in terms of the observed
lensing based profiles.  The distortion profiles measured are amongst
the most accurate constructed to date and the great depth of the
Subaru imaging permits magnification profiles to be 
established with unprecedented detail using the background red galaxy
counts. The magnification and distortion profiles are fitted well by
the general NFW profile for CDM dominated halos, which continuously
steepens with radius, but the profiles derived are more concentrated
than predicted by $\Lambda$CDM based simulations for which clusters 
of $\sim 10^{15}M_{\odot}$, form late ($z<0.5$) with low
concentrations, reflecting the low mean cosmological mass density today.

Tension with the standard $\Lambda$CDM model is also indicated by the
observed Einstein radii of very massive clusters, including those
studied here, which lie well beyond the predicted range derived from
massive halos generated in $\Lambda$CDM based simulations (Broadhurst
\& Barkana 2008). In addition, a significant discrepancy is also
claimed for X-ray derived cluster concentrations, with respect to the
$\Lambda$CDM based on WMAP5 parameters (Duffy et al. 2008). The
combined X-ray and lensing based concentrations compiled by Comerford
and Natarajan (2008) also exceeds somewhat the standard expectation,
despite including X-ray based concentrations that are found to be
systematically lower than for lensing (Comerford and Natarajan 2008).
The mass-concentration relation of Mandelbaum etal (2008) is claimed to
be more consistent with $\Lambda$CDM. Here allowance is made for the
smoothing of the central lensing signal when stacking many SDSS
clusters, relying on the brightest cluster galaxy to define the
center of mass. Cluster members misclassified as background
galaxies when estimating photometric redshifts will also act to
reduce the derived concentrations, diluting the central lensing
signal.

Without resorting to radical proposals regarding the nature of DM, our
findings imply the central region of clusters collapsed earlier than
expected. Assuming the simple redshift relation $c_{\rm vir}(z)=c_0
(1+z)$ (e.g., Wechsler et al. 2006), relating central cluster densities
to the cosmological mean density, then the formation of clusters with
$c_{\rm vir}(1+z)=10-15$ corresponds to $z \geq 1$, significantly
earlier than in the standard $\Lambda$CDM, for which clusters form today
with $c_0 \sim 5.6$.

The presence of massive clusters at high redshift ($z\sim 2$), and the
old ages of their member galaxies (Zirm et al. 2008; Blakeslee et al.
2003), may also imply clusters collapsed at relatively early
times(Mathis, Diego, \& Silk 2004), for which accelerated growth factors
have been proposed, adopting a generalised equation of state (Sadeh \&
Rephaeli 2008). Alternatively, since clusters correspond to rare
density maxima, then any non-Gaussianity in the early fluctuation
spectrum may advance cluster formation (Matthis, Diego, \& Silk 2004;
Sadeh, Rephaeli, \& Silk 2007). Our results present a challenge to
some of these models, since earlier cluster formation must not also
enhance significantly the abundance of clusters.


\acknowledgements

This work is in part supported by the Israeli Science
Foundation and the National Science Council of Taiwan under the grant
NSC95-2112-M-001-074-MY2 and by the US Department of Energy, contract 
DE-AC02-76SF00515.


\clearpage


\begin{figure}[t]
\begin{center}
\resizebox{16cm}{!}{\includegraphics{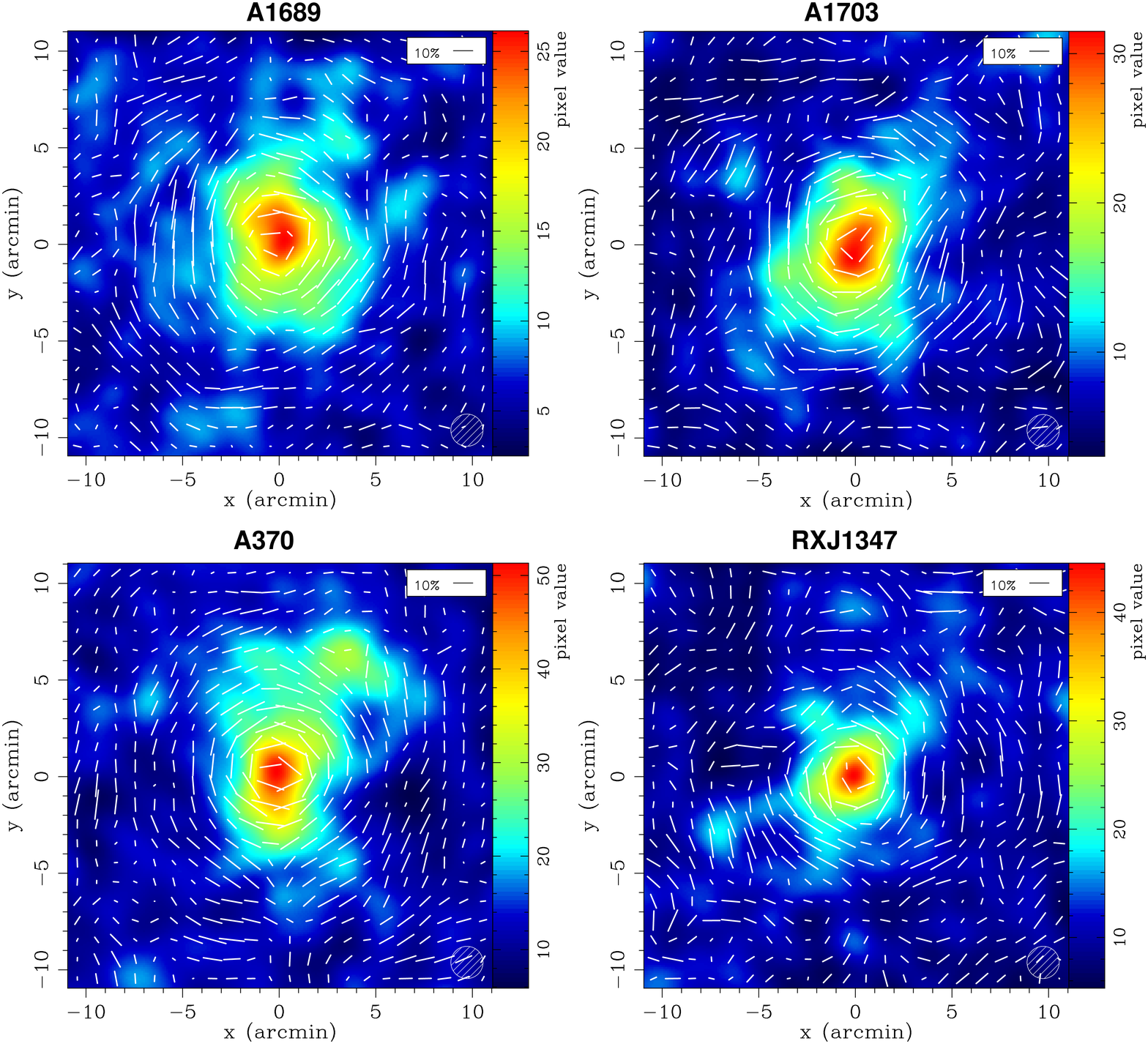}}
\end{center}
\caption{Maps of the surface number-density distribution of
color-selected cluster member galaxies, with the gravitational shear
of background galaxies overlayed; $10\%$ ellipticity is indicated top
right, and the resolution of the distortion map is shown bottom right.
In each case a single concentration of galaxies is visible, around
which a coherent tangential pattern is centered, with little
significant substructure.
}
\label{fig:shear}
\end{figure}

\begin{figure}[tbh]
 \begin{center}
 \hglue0.0cm{\includegraphics[angle=0,width=8.cm]{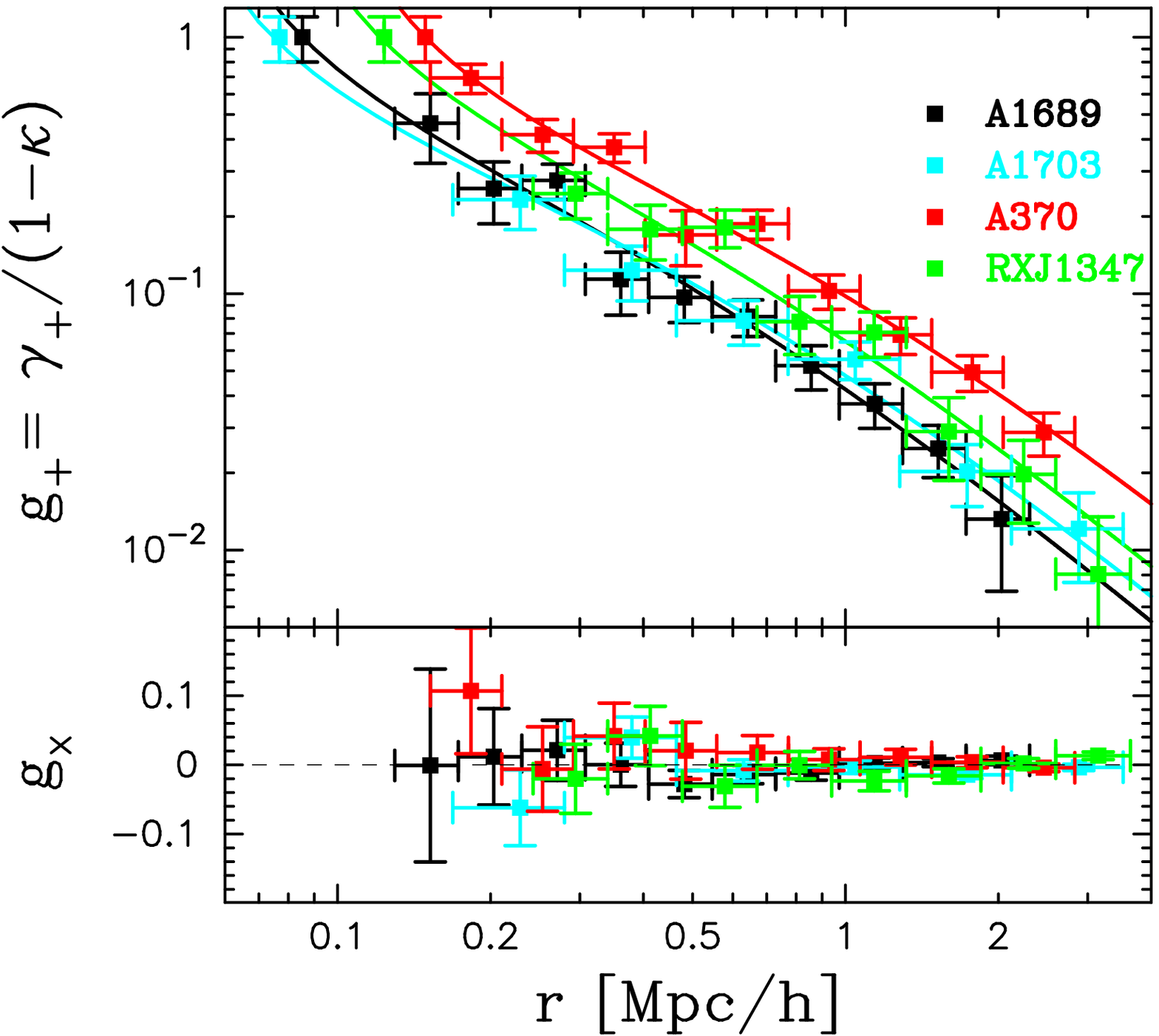}}
 \hglue0.5cm{\includegraphics[angle=0,width=8.cm]{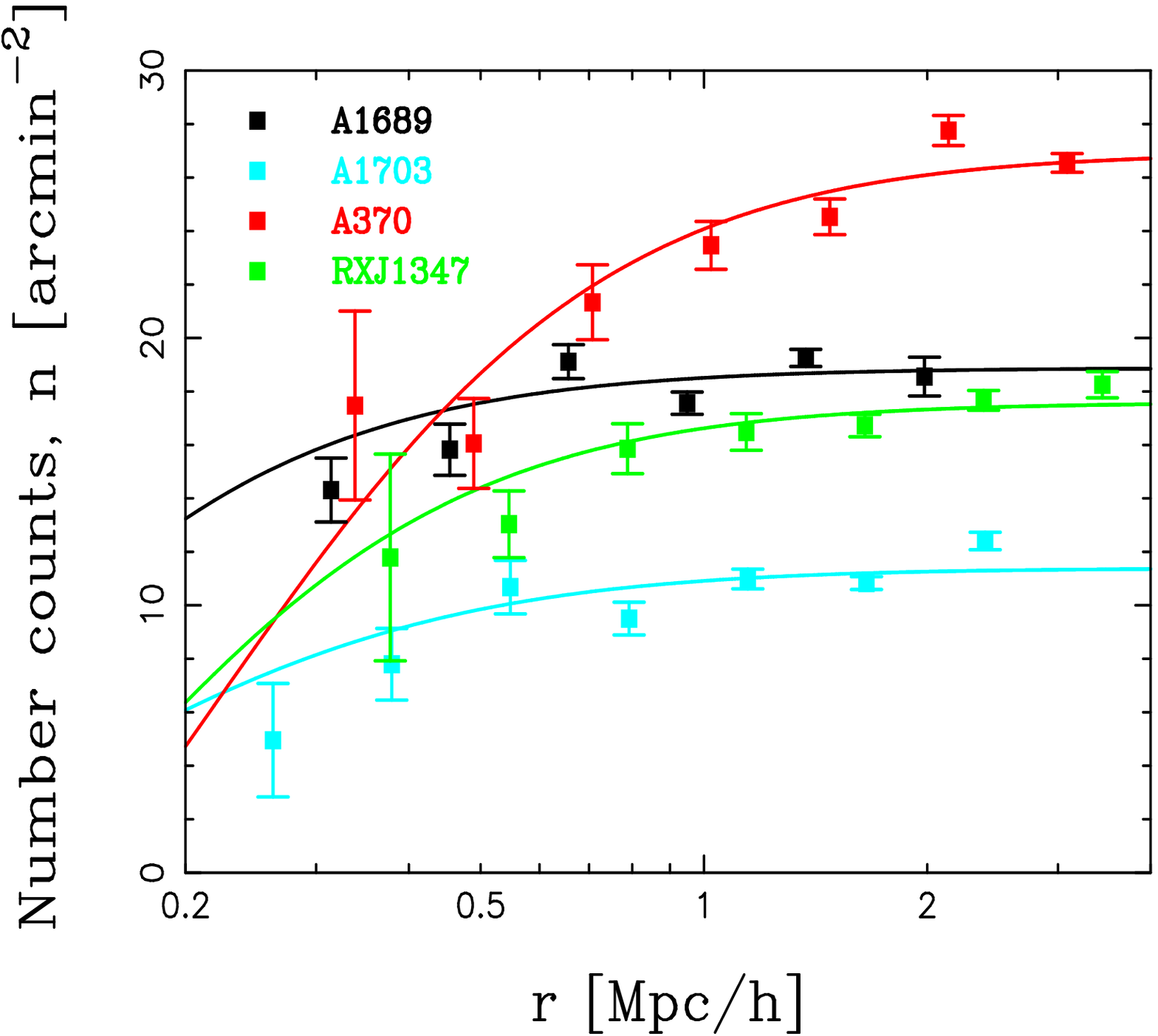}}
   \caption {({\it Left}): Tangential distortion profiles of the 4
     clusters based on the combined red and blue background samples
     and compared with the best-fitting NFW model. The Einstein-radius
     determined from multiply-lensed images is indicated, marking the
     point of maximum distortion, $g_+=1$. Shown below is the rotated
     shear, $g_x$, demonstrating no obvious non-tangential
     distortion. ({\it Right}): Profiles of background red galaxy
     counts, whose intrinsic slope is relatively shallow, so lens
     magnification reduces the observed numbers towards the center.
     Over-plotted are normalised NFW models derived from the
     distortion profiles in the left panel, demonstrating consistency
     between these two independent lensing observables; note A370 is
     shifted upward 40\% for clarity. The count uncertainty is
     obtained by sub-dividing each annulus into equal area cells, with
     a tail of $>2\sigma$ cells excluded to remove inherent small
     scale clustering of the background. Small areas around each
     bright object ($m_R<21$) are excluded, including cluster members,
     interior to 3 times the isophotal radius, where the detection of
     faint galaxies is significantly compromised (Broadhurst et al.
     2005b; Umetsu \& Broadhurst 2008).  }
   \label{one}
 \end{center}
 \end{figure}

\begin{figure}[t]
\begin{center}
  \resizebox{14cm}{!}{\includegraphics{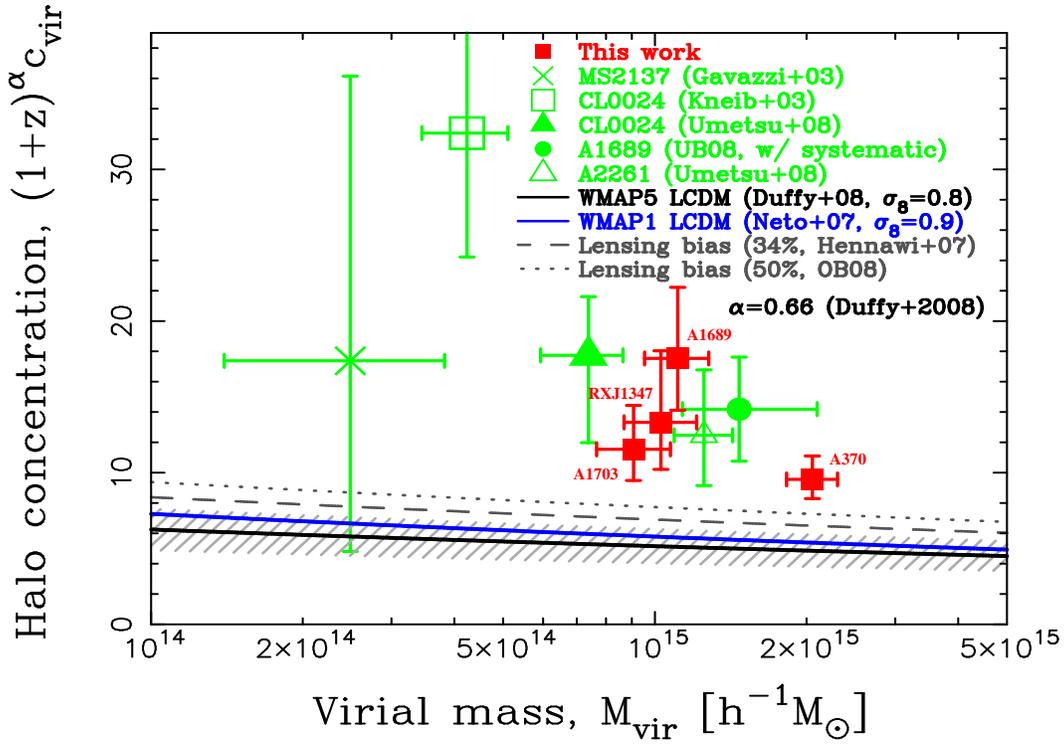}}
\end{center}
\caption{ Comparison of observations with the $\Lambda$CDM model, based on
$N$-body simulations for the $c_{\rm vir}-M_{\rm vir}$ relation,
derived at $z=0$. The predictions of Duffy et al. (2008)
($\sigma_8=0.8$, WMAP5) and Neto et al. (2007)
($\sigma_8=0.9$, WMAP1) are shown as solid curves, with
$1\sigma$ uncertainty (from Neto et al. 2007) indicated by the hatched
area. Also shown is the level of selection and orientation bias for
projected masses based on $\Lambda$CDM (dashed curves, see text). The
data points are all derived from lensing alone, and are multiplied by
$(1+z)^{0.66}$ at the cluster redshift, for consistency with the
evolution of $c_{\rm vir}(M_{\rm vir})$ derived from $\Lambda$CDM
simulations by Duffy et al. 2008. Note, a significant correction for
substructure is required for Cl0024+16 (Umetsu et al., in
preparation). Clearly the data lie well above the predicted relation even
when possible sources of bias are considered.}
\label{fig:magbias}
\end{figure}

\end{document}